\documentclass{midl} % Include author names
\newcommand\figwidth{0.1035}
\newcommand\figgap{-3.25}
\newcommand{\Revision}[1]{\textcolor{black}{#1}}

\usepackage{mwe} % to get dummy images
\usepackage{bm}
\usepackage{booktabs}
\usepackage{siunitx}
\usepackage{nth}
\usepackage{multirow}
\usepackage{makecell}
\usepackage{rotating}
\title[MT-UNet]{\Revision{Multi-task UNet: Jointly Boosting Saliency Prediction and Disease Classification on Chest X-ray Images}}

\midlauthor{\Name{Hongzhi Zhu\nametag{$^{1}$}} \Email{hzhu@ece.ubc.ca}
\AND
\Name{Robert Rohling\nametag{$^{1,2,3}$}} \Email{rohling@ece.ubc.ca}
\AND
\Name{Septimiu Salcudean\nametag{$^{1,2}$}} \Email{tims@ece.ubc.ca}\\
\addr $^{1}$ School of Biomedical Engineering, University of British Columbia, Vancouver, Canada\\
\addr $^{2}$ Department of Electrical and Computer Engineering, University of British Columbia, Vancouver, Canada\\
\addr $^{3}$ Department of Mechanical Engineering, University of British Columbia, Vancouver, Canada
}

\begin{document}

\maketitle

\begin{abstract}
Human visual attention has recently shown its distinct capability in boosting machine learning models. However, studies that aim to facilitate medical tasks with human visual attention are still scarce. To support the use of visual attention, this paper describes a novel deep learning model for visual saliency prediction on chest X-ray (CXR) images. To cope with data deficiency, we exploit the multi-task learning method and tackles disease classification on CXR simultaneously. For a more robust training process, we propose a further optimized multi-task learning scheme to better handle model overfitting. Experiments show our proposed deep learning model with our new learning scheme can outperform existing methods dedicated either for saliency prediction or image classification. The code used in this paper is available at \url{https://github.com/hz-zhu/MT-UNet}.
\end{abstract}


\section{Introduction}\label{sec:Introduction}
Recent work in machine learning and computer vision have demonstrated advantages of integrating human attention with artificial neural network models, as studies show that many machine vision tasks, i.e., image segmentation, image captioning, object recognition, etc., can benefit from adding human visual attention \cite{liu2018visual}.
\par
%Visual attention is the ability inherited in biological visual systems to selectively recognize regions or features on scenes relevant to a specific task \cite{borji2012quantitative}. Two major components shape the visual attention: the ``top-down'' component based on the task and the visual system, and the ``bottom-up'' component where certain features on scenes pertinent to the task distinguish themselves from the rest \cite{borji2012quantitative,cong2018review}. Early research on saliency prediction focuses more on the uncovering of the ``bottom-up'' component \cite{borji2012quantitative}, while more recent attempts, empowered by machine learning, start to consider both components, and thus the terms, saliency prediction and visual attention prediction, are used interchangeably \cite{sun2021visual}. In this paper, we use the term saliency prediction as the prediction of human visual attention. With the establishment of several benchmark datasets, data driven approaches demonstrated major advancements in saliency prediction (review in \citet{borji2019saliency} and \citet{wang2019revisiting}). However, saliency prediction for natural scenes is the primary focus, and little work has been done in the medical domain. Hence, we intend to fill the gap and study the saliency prediction for examining chest X-ray (CXR) images, one of the most common radiology tasks worldwide.
\Revision{Visual attention is the ability inherited in biological visual systems to selectively recognize regions or features on scenes relevant to a specific task \cite{borji2012quantitative}, where ``bottom-up'' attention (also called exogenous attention) focuses on physical properties in the visual input that are salient and distinguishable, and ``top-down'' attention (also called endogenous attention) generally refers to mental strategies adopted by the visual systems to accomplish the intended visual tasks \cite{paneri2017top}. Early research on saliency prediction aims to understand attentions triggered by visual features and patterns, and thus ``bottom-up'' attention is the research focus \cite{borji2012quantitative}. More recent attempts, empowered by interdisciplinary efforts, start to study both ``bottom-up'' and ``top-down'' attentions, and therefore the terms, saliency prediction and visual attention prediction, are used interchangeably \cite{sun2021visual}. In this paper, we use the term saliency prediction as the prediction of human visual attentions allocations when viewing 2D images, containing both ``bottom-up'' and ``top-down'' attentions. 2D heatmap is usually used to represent human visual attention distribution. Note that saliency prediction studied in this paper is different from neural network's saliency/attention which can be visualized through class activation mapping (CAM) by \citet{zhou2016learning} and other methods \cite{simonyan2013deep, fu2019multicam,selvaraju2016grad}. With the establishment of several benchmark datasets, data driven approaches demonstrated major advancements in saliency prediction (review in \citet{borji2019saliency} and \citet{wang2019revisiting}). However, saliency prediction for natural scenes is the primary focus, and more needs to be done in the medical domain. Hence, we intend to study the saliency prediction for examining chest X-ray (CXR) images, one of the most common radiology tasks worldwide.}

\par
CXR imaging is commonly used for the diagnosis of cardio and/or respiratory abnormalities; it is capable of identifying multiple conditions through a single shot, i.e., COVID-19, pneumonia, heart enlargement, etc. \cite{ccalli2021deep}. There exists multiple public CXR datasets \cite{irvin2019chexpert,wang2017chestx}. However, the creation of large comprehensive medical datasets is labour intensive, and requires significant medical resources which are usually scarce \cite{castro2020causality}. Consequently, medical datasets are rarely as abundant as that for non-medical fields. Thus, machine learning approaches applied on medical datasets need to address the problem of data scarcity. In this paper, we exploit the multi-task learning for solution.
\par
Multi-task learning is known for its inductive transfer characteristics that can drive strong representation learning and generalization of each component task \cite{caruana1997multitask}. Therefore, multi-task learning methods partially alleviates some of the major shortcomings in deep learning, i.e., high demands for data sufficiency and heavy computation loads \cite{crawshaw2020multi}. However, to apply multi-task learning methods successfully, challenges still exist, which can be the proper selection of component tasks, the architecture of the network, the optimization of the training schemes and many others \cite{zhang2021survey,crawshaw2020multi}. This paper investigates the proper configuration of a multi-task learning model that can tackle visual saliency prediction and image classification simultaneously.
\par
The main contributions of this paper are: 1) development of a new deep convolutional neural network (DCNN) architecture for CXR image saliency prediction and classification based on UNet \cite{ronneberger2015u}, and 2) proposal of an optimized multi-task learning scheme that handles overfitting. Our method aims to outperform the state-of-the-art networks dedicated either for saliency prediction or image classification.

\section{Background}
\subsection{Saliency prediction with deep learning}
%Two types of saliency prediction exist in the literature: static saliency prediction over a still scene or image, and dynamic saliency over a motion scene or video. 
DCNN is the leading machine learning method applied to saliency prediction \cite{pan2016shallow, kummerer2016deepgaze, jia2020eml, kroner2020contextual}. Besides, transfer learning with pre-trained networks was observed to boost the performance of saliency prediction \cite{oyama2017fully, kummerer2016deepgaze, oyama2018influence}.
A majority of DCNN approaches are for natural scene saliency prediction, and so far, only a few studied the saliency prediction for medical images. By \citet{cai2018multi}, the generative adversarial network is used to predict expert sonographer's saliency when performing standard fetal head plane detection on ultrasound (US) images. However, the saliency prediction is used as a secondary task to assist the primary detection task, and thus, the saliency prediction performance failed to outperform benchmark prediction methods in several key metrics. Similarly, by \citet{karargyris2021creation}, as a proof-of-concept study, the gaze data is used as an auxiliary task for CXR image classification, and the performance of saliency prediction is not reported in the study.

\subsection{CXR image classification with deep learning}
Public datasets for CXR images enabled data driven approaches for automatic image analysis and diagnosis \cite{serte2020deep,li2020accuracy}. Advancements in standardized image classification networks, i.e., ResNet \cite{he2016deep}, DenseNet \cite{huang2017densely}, and EfficientNet \cite{tan2019efficientnet}, facilitate CXR image classification. Yet, CXR image classification remains challenging, as CXR images are noisy, and may contain subtle features that are difficult to recognize even by experts \cite{ccalli2021deep,khan2021intelligent}. 

\section{Multi-task Learning Method}
As stated in Section \ref{sec:Introduction}, component task selection, network architecture design, and training scheme are key factors for multi-task learning. We select the classification task together with the saliency prediction based on the fact that attention patterns are task specific \cite{karessli2017gaze}. \Revision{Radiologists are likely to exhibit distinguishable visual behaviors when different patient conditions are shown on CXR images \cite{mclaughlin2017computing}.} This section introduces our multi-task UNet (MT-UNet) architecture, and derives a better multi-task training scheme for saliency prediction and image classification.

\begin{figure}[t]
\floatconts
  {fig:MTL_UNet}
  {\caption{MT-UNet architecture. The solid blocks represent 3D tensors, $\mathbf{R}^{F\times H\times W}$, where $F$, $H$, and $W$ denote feature (channel), height and width dimensions, respectively. The solid circles represent 1D tensors. Arrows denote operations to the tensors. Numbers above some of the solid blocks stand for the number features in tensors.}}
  {\includegraphics[width=0.80\linewidth]{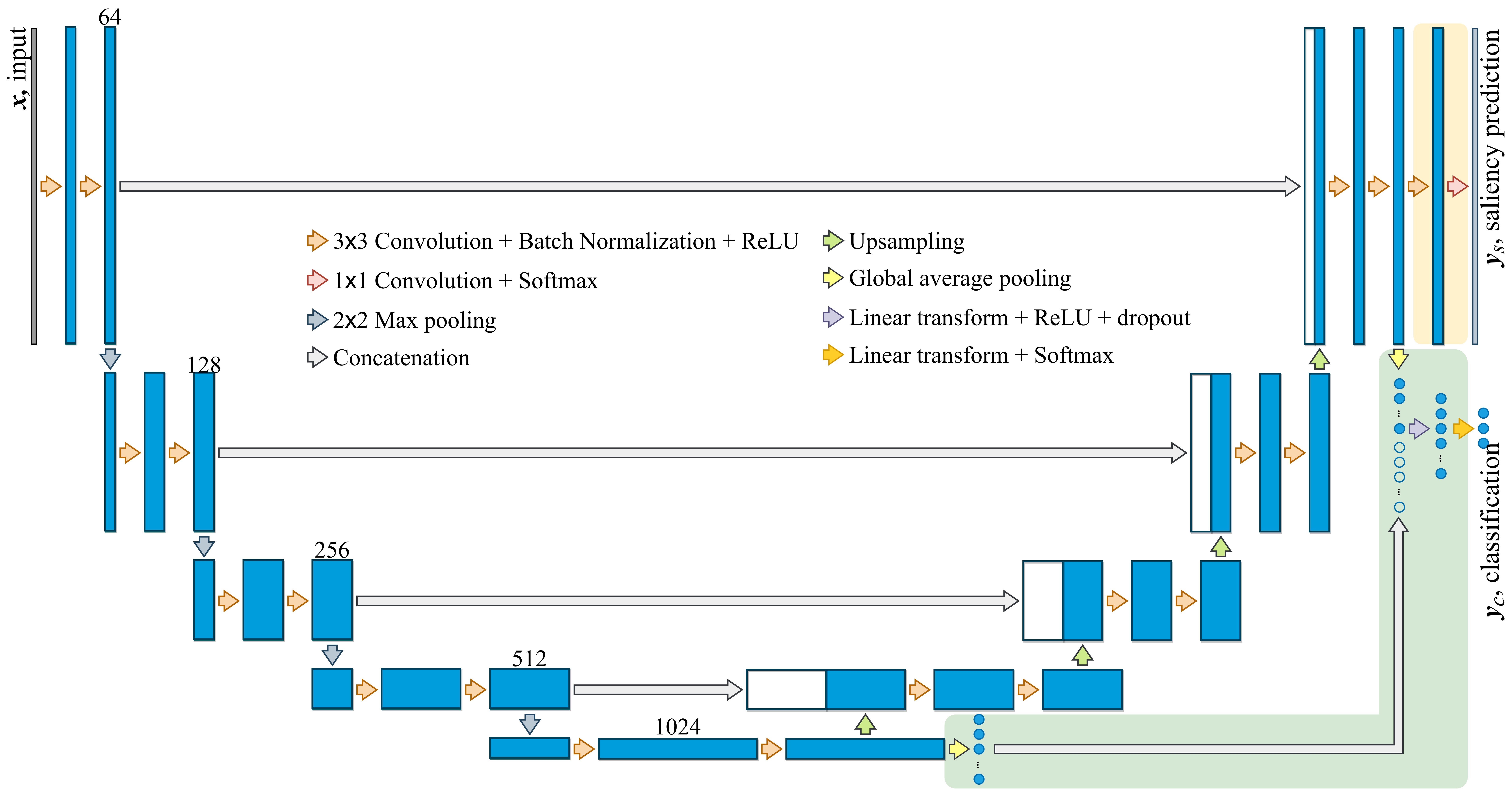}}
\end{figure}

\subsection{Multi-task UNet}
\label{sec:Multi-task UNet}
\Revision{\figureref{fig:MTL_UNet} shows the architecture of the proposed MT-UNet. The network takes CXR images, $\bm{x}\in\mathbf{R}^{1\times H\times W}$, where $H$ and $W$ are image dimensions, as input, and produces two outputs, predicted saliency $\bm{y}_s\in\mathbf{R}^{1\times H\times W}$, and predicted classification $\bm{y}_c\in\mathbf{R}^{C}$, where $C$ is the number of classes. 
As the ground truth for $\bm{y}_s$ is human visual attention distribution, represented as a 2D matrix whose elements are non-negative and sum to $1$, $\bm{y}_s$ is normalized by Softmax before output from MT-UNet. Softmax is also applied to $\bm{y}_c$ before output so that the classification outcome can be interpreted as class probability.} For the simplicity of notation, batch dimensions are neglected.
\par
%The standard UNet architecture is visible from \figureref{fig:MTL_UNet}. As a well-known image-to-image deep learning model, the UNet structure has been adopted for various tasks. For example, the UNet is appended with additional structures for visual scene understanding \cite{jha2020mt}, and the features from the middle of the UNet are extracted for image classification tasks \cite{karargyris2021creation}.
\Revision{The proposed MT-UNet is derived from standard UNet architecture \cite{ronneberger2015u}. As a well-known image-to-image deep learning model, the UNet structure has been adopted for various tasks. For example, the UNet is appended with additional structures for visual scene understanding \cite{jha2020mt}, the features from the bottleneck (middle of the UNet) are extracted for image classification tasks \cite{karargyris2021creation}, and by combining UNet with Pyramid Net \cite{lin2017feature}, features at different depth are aggregated for enhanced segmentation \cite{moradi2019mfp}. 
What's more, the encoder-decoder structure of UNet is utilized for multi-task learning, where the encoder structure is used to learn representative features, along with designated decoder structures or classification heads for image reconstruction, segmentation, and/or classification \cite{zhou2021multi, amyar2020multi}.
In our design, we apply classification heads (shaded in light green in \figureref{fig:MTL_UNet}), which are added not only to the bottleneck but also the ending part of the UNet architecture.} This additional classification specific structures aggregates middle and higher-level features for classification, exploiting features learnt at different depths. The attention heads perform global average pooling operations to the 4D tensors, followed by concatenation, and two linear transforms (dense layers) with dropout (rate=$25\%$) in the middle to produce classification outcomes. 
%The MT-UNet belongs to the hard parameter sharing structure in multi-task learning, where different tasks share the same trainable parameters before branched out to each tasks' specific parameters \cite{vandenhende2021multi}. In \figureref{fig:MTL_UNet}, we use yellow and green shades to denote network structures dedicated for saliency prediction and classification, respectively.
%Usually, multi-task learning using hard parameter sharing schemes has drawbacks of introducing too many additional parameters to the network \cite{vandenhende2021multi}. Therefore, in our design, we wish to avoid this drawback by minimizing task specific structures.
\Revision{The MT-UNet belongs to the hard parameter sharing structure in multi-task learning, where different tasks share the same trainable parameters before branched out to each tasks' specific parameters \cite{vandenhende2021multi}. Having more trainable parameters in task specific structures may improve the performance for that task at a cost of introducing additional parameters and increasing computational load \cite{crawshaw2020multi, vandenhende2021multi}. In our design, we wish to avoid heavy structures with lots of task specific parameters, and therefore, task specific structures are minimized. In \figureref{fig:MTL_UNet}, we use yellow and green shades to denote network structures dedicated for saliency prediction and classification, respectively.}

\subsection{Multi-task Training Scheme}
Balancing the losses between tasks in a multi-task training process has a direct impact on the training outcome \cite{vandenhende2021multi}. There exist multi-task training schemes \cite{kendall2018multi, chen2018gradnorm, guo2018dynamic, sener2018multi}, and among which, we adopt the uncertainty based balancing scheme \cite{kendall2018multi} with the modification proposed in \cite{liebel2018auxiliary}. Hence, the loss function is:
\begin{equation}
\label{eq:loss_1}
    \mathcal{\bm{L}} = \frac{1}{\sigma_s^2}L_s+\frac{1}{\sigma_c^2}L_c+\ln(\sigma_s+1)+\ln(\sigma_c+1)
\end{equation}
where $L_s$ and $L_c$ are loss values for $\bm{y}_s$ and $\bm{y}_c$, respectively; $\sigma_s>0$ and $\sigma_c>0$ are trainable scalars estimating the uncertainty of $L_s$ and $L_c$, respectively; $\sigma_s$ and $\sigma_c$ are initialized to $1$; $\ln(\sigma_s+1)$ and $\ln(\sigma_c+1)$ are regularizing terms to avoid arbitrary decrease of $\sigma_s$ and $\sigma_c$.
\Revision{With \equationref{eq:loss_1}, we know that $\sigma$ values can dynamically weigh losses of different amplitudes during training, and loss with low uncertainty (small $\sigma$ value) is prioritized in the training process. $\mathcal{\bm{L}}>0$.}
%As stated in \sectionref{sec:Multi-task UNet}, $\bm{y}_s$ and $\bm{y}_c$ are normalized with Softmax to represent distributions, so the loss function for $\bm{y}_s$ and $\bm{y}_c$ are Kullback-Leibler divergence (KLD) loss and cross-entropy loss, respectively.
\Revision{Given $\bm{y}_s$ and $\bm{y}_c$ with their ground truth $\bar{\bm{y}}_s$ and $\bar{\bm{y}}_c$, respectively, the loss functions are:}
\begin{equation}
    L_s = H(\bar{\bm{y}}_s, {\bm{y}}_s)-H(\bar{\bm{y}}_s),
    \label{eq:L_s}
\end{equation}
\begin{equation}
    L_c = H(\bar{\bm{y}}_c, {\bm{y}}_c) \quad \quad \quad \quad
    \label{eq:L_c}
\end{equation}
\Revision{where $H(Q,R)=-\Sigma_{i}^nQ_i\ln(R_i)$ stands for cross entropy of two discrete distributions $Q$ and $R$, both with $n$ elements; $H(Q)=H(Q,Q)$ stands for the entropy, or self cross entropy, of discrete distribution $Q$. $L_s$ is the Kullback-Leibler divergence (KLD) loss, and $L_c$ is the cross-entropy loss.}
%Since the term $-H(\bar{\bm{y}}_s)$ in $L_s$ is a constant and does not generate gradient when updating parameters, in \equationref{eq:loss_1}, we follow the method derived in \cite{kendall2018multi} to scale cross entropy loss with , and use $\frac{1}{\sigma_s^2}$} to scale KLD loss ($L_s$)
\Revision{
By observing \equationref{eq:L_s} and \equationref{eq:L_c}, we know that only the cross entropy terms, $H(\cdot, \cdot)$, generate gradient when updating network parameters, as the term $-H(\bar{\bm{y}}_s)$ in $L_s$ is a constant and has zero gradient. Therefore, we extend the method in \cite{kendall2018multi}, and use $\frac{1}{\sigma^2}$ to scale a KLD loss ($L_s$) as that for a cross-entropy loss ($L_c$).}
\par
\Revision{
Although the training scheme in \equationref{eq:loss_1} yields many successful applications, overfitting for multi-task networks still can jeopardize the training process, especially for small datasets \cite{wang2020makes}. Multiple factors can cause overfitting, among witch, learning rate, $r>0$, shows the most significant impact \cite{li2019research}. Also, $r$ generally has significant influences on the training outcome \cite{smith2018disciplined}, making it one of the most important hyper-parameters for a training process.
When training MT-UNet, $r$ is moderated by several factors. The first factor is the use of an optimizer. Many optimizers, i.e., Adam \cite{kingma2014adam} and RMSProp \cite{tieleman2012lecture}, deploy the momentum mechanism or its variants, which can adaptively adjust the effective learning rate, $r_e$, during training. As a learning rate scheduler is often used for more efficient training, it is the second factor to influence $r$. The influence of $r$ from a learning rate scheduler can be adaptive, i.e., reduce learning rate on plateau (RLRP), or more arbitrary, i.e., cosine annealing with warm restarts \cite{loshchilov2016sgdr}. By observing \equationref{eq:loss_1}, we know that an uncertainly estimator $\sigma$ for a loss $L$ also serves as a learning rate adaptor for $L$, which is the third factor. More specifically, given a loss value $L$ with learning rate $r$, the effective learning rate for parameters with a scaled loss value $\frac{L}{\sigma^2}$ is $\frac{r}{\sigma^2}$.}
\par
\begin{figure}[thbp]
    \centering
    \subfigure[Losses]{
	   \includegraphics[width=0.35\textwidth]{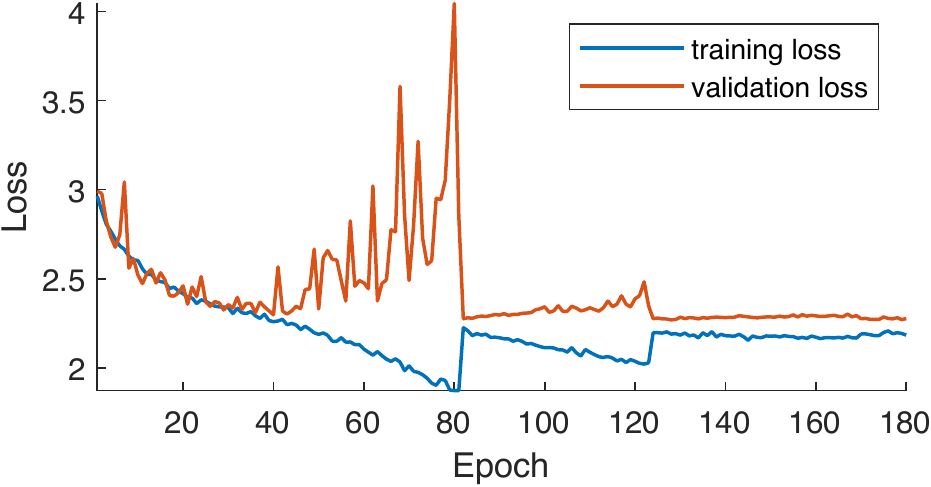}
	   \label{fig:training_losses}
	   }
     \hfill
    \subfigure[$\sigma$ values]{
	   \includegraphics[width=0.35\textwidth]{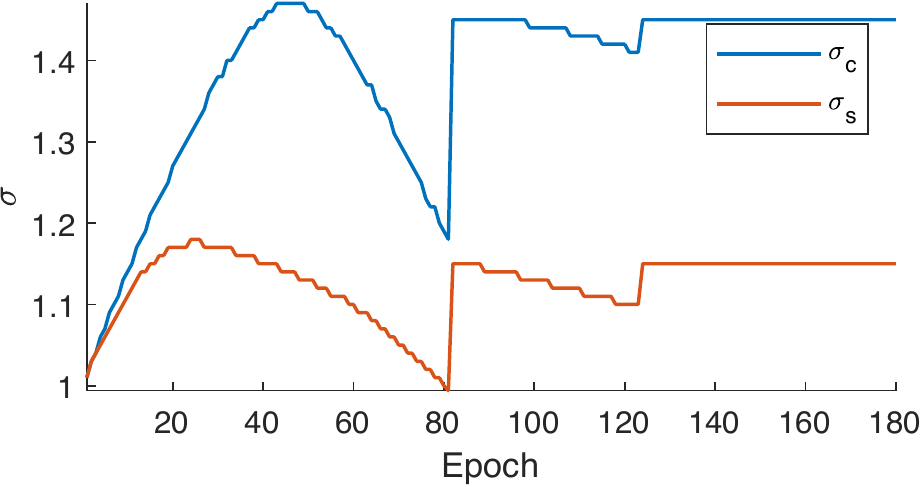}
	   \label{fig:training_sigmas}
	   }
\caption{Training process visualization with \equationref{eq:loss_1}}
\label{fig:training}
\end{figure}

\Revision{
Decreasing $r$ upon overfitting can alleviate its effects \cite{smith2018disciplined, duffner2007online}, but \equationref{eq:loss_1} leads to increased learning rate upon overfitting, further worsening the training process. This happens because training loss decreases when overfitting occurs, reducing its variance at the same time. Thus, $\sigma$ decreases accordingly, which increases the effective learning rate, thus creating a vicious circle of overfitting. More detailed mathematical derivation is presented in Appendix \ref{app:math}. This phenomenon can be observed in \figureref{fig:training}, where changes of losses and $\sigma$ values during a training process following \equationref{eq:loss_1} are presented.
We can see from \figureref{fig:training_losses}, at epoch $40$, after an initial decrease in both the training and validation losses, the training loss start to acceleratedly decrease while the validation loss start to amplify, which is a vicious circle of overfitting. A RLRP scheduler can halt the vicious circle by resetting the model parameters to a former epoch and reducing $r$. Yet, even with reduced $r$, a vicious circle of overfitting can remerge in later epochs.
}

To alleviate overfitting, we propose the use of the following equations to replace \equationref{eq:loss_1}: 
\begin{equation}
    \mathcal{\bm{L}} = \frac{1}{\sigma_s^2}L_s+L_c+\ln(\sigma_s+1),
    \label{eq:loss_2}
\end{equation}
\begin{equation}
    \mathcal{\bm{L}} = L_s+\frac{1}{\sigma_c^2}L_c+\ln(\sigma_c+1).
    \label{eq:loss_3}
\end{equation}
The essence of \equationref{eq:loss_2,eq:loss_3} is to fix the uncertainty term for one loss in \equationref{eq:loss_1} to $1$, so that the flexibility in changing effective learning rate is reduced. With the uncertainty term fixed for one component loss, \equationref{eq:loss_2,eq:loss_3} demonstrate the ability to alleviate overfitting and stabilize the training processing. It is worth noting that \equationref{eq:loss_2,eq:loss_3} cannot be used interchangeably. 
\Revision{We need to test both equations to check which can achieve better performances, as depending on the dataset and training process, overfitting can occur of different severity in all component tasks.}
In this study, training process with \equationref{eq:loss_3} achieves the best performance. Ablation study of this method is presented in \sectionref{sec:Experiment and Result}.

\section{Dataset and Evaluation Methods}
We use the ``chest X-ray dataset with eye-tracking and report dictation'' \cite{karargyris2021creation} shared via PhysioNet \cite{moody2000physionet} in this study. The dataset was derived from the MIMIC-CXR dataset \cite{johnson2019mimic, johnson2019mimic2} with additional gaze tracking and dictation from an expert radiologist. $1083$ CXR images are included in the dataset, and accompanying each image, there are tracked gaze data; a diagnostic label (either normal, pneumonia, or enlarged heart); segmentation of lungs, mediastinum, and aortic knob; and radiologist's audio with dictation. The CXR images in the dataset are in resolutions of various sizes, i.e., $3056\times2044$, and we down sample and/or pad each image to $640\times416$. A GP3 gaze tracker by Gazepoint (Vancouver, Canada) was used for the collection of gaze data. The tracker has an accuracy of around \SI{1}{\degree} of visual angle, and has a \SI{60}{\Hz} sampling rate \cite{zhu2019novel}.
\par
Several metrics have been used for the evaluation of saliency prediction performances, and they can be classified into location-based metrics and distribution-based metrics \cite{bylinskii2018different}. Due to the tracking inaccuracy of the GP3 gaze tracker, location-based metrics is not suited for this study. Therefore, in this paper, we follow suggestions in \cite{bylinskii2018different} and use KLD for performance evaluation. We also include histogram similarity (HS), and Pearson's correlation coefficient (PCC) for reference purposes.
\Revision{For the evaluation of classification performances, we use the area under curve (AUC) metrics for multi-class classifications \cite{hand2001simple, fawcett2006introduction}, and the classification accuracy (ACC) metrics. We also include the AUC metrics for each class: normal, enlarged heart, and pneumonia, denoted as AUC-Y1, AUC-Y2, and AUC-Y3, respectively.}
\Revision{
In this paper, all metrics values are presented as median statistics followed by standard deviations behind the $\pm$ sign. Metrics with up-pointing arrow $\uparrow$ indicates greater values reflect better performances, and vise versa. Best metrics are emboldened.}
\par

\section{Experiments and Result}
\label{sec:Experiment and Result}
%To validate our MT-UNet architecture and the optimized training scheme, comparisons are performed with multiple benchmark architectures for CXR image classification and saliency prediction. Ablation study is also performed to confirm the improvement of our optimized training scheme and architecture.

\subsection{Benchmark comparison}
\label{sec:Benchmark comparison}
In this subsection, we compare the performance of MT-UNet, with benchmark networks for CXR image classification and saliency prediction. Detailed training settings are presented in Appendix \ref{app:Training settings}.
\par
For CXR image classification, the benchmark networks are chosen from the top performing networks for CXR image classification examined in \cite{el2021automated}, which are ResNet50 \cite{he2016deep} and Inception-ResNet v2 (abbreviated as IRNetV2 in this paper) \cite{szegedy2017inception}. Following \citet{karargyris2021creation}, we also include a state-of-the-art general purpose classification network: EfficientNetV2-S (abbreviated as EffNetV2-S) \cite{tan2021efficientnetv2} for comparison. \Revision{For completeness, classification using standard UNet with additional classification head (denoted as UNetC) is included.} Results are presented in \tableref{tab:compare_classification}, and We can see that MT-UNet outperforms the other classification networks.
\par
For CXR image saliency prediction, comparison was conducted with $3$ state-of-the-art saliency prediction models, which are SimpleNet \cite{reddy2020tidying}, MSINet \cite{kroner2020contextual} and VGGSSM \cite{cao2020aggregated}. \Revision{Saliency prediction using standard UNet (denoted as UNetS) is also included for reference.} Table \ref{tab:compare_saliency_result} shows the result, where MT-UNet outperforms the rest. \Revision{Visual comparisons for saliency prediction results are presented through \tableref{tab:cam visual} in Appendix \ref{app:Performance evaluation}.}

\begin{table}[thbp]
\centering
\begin{tabular}{@{}c|ccccc@{}}
\toprule
Metrics &MT-UNet &UNetC &EffNetv2-S &IRNetv2 &ResNet50 \\ \midrule
ACC $\uparrow$ &$\bm{0.670}\pm0.018$ &$0.593\pm0.009$ &$0.640\pm0.037$ &$0.640\pm0.017$ &$0.613\pm0.013$ \\ 
AUC $\uparrow$ &$\bm{0.843}\pm0.012$ &$0.780\pm0.006$ &$0.826\pm0.015$ &$0.824\pm0.014$ &$0.816\pm0.010$ \\
AUC-Y1 $\uparrow$ &$\bm{0.864}\pm0.014$ &$0.841\pm0.007$ &$0.852\pm0.013$ &$0.862\pm0.016$ &$0.845\pm0.015$ \\
AUC-Y2 $\uparrow$ &$\bm{0.912}\pm0.008$ &$0.840\pm0.003$ &$0.901\pm0.015$ &$0.897\pm0.011$ &$0.896\pm0.015$ \\
AUC-Y3 $\uparrow$ &$\bm{0.711}\pm0.027$ &$0.597\pm0.018$ &$0.653\pm0.017$ &$0.633\pm0.036$ &$0.622\pm0.022$ \\
\bottomrule
\end{tabular}
\caption{\Revision{Performance comparison between classification models.}}
\label{tab:compare_classification}
\end{table}

\begin{table}[thbp]
\centering
\begin{tabular}{@{}c|ccccc@{}}
\toprule
Metrics &MT-UNet &UNetS &SimpleNet &MSINet &VGGSSM \\ \midrule
KLD $\downarrow$ &$\bm{0.726}\pm0.004$ &$0.750\pm0.002$ &$0.758\pm0.009$ &$0.748\pm0.003$ &$0.743\pm0.007$ \\
PCC $\uparrow$ &$\bm{0.569}\pm0.004$ &$0.552\pm0.002$ &$0.545\pm0.008$ &$0.557\pm0.002$ &$0.561\pm0.005$ \\
HS $\uparrow$ &$\bm{0.548}\pm0.001$ &$0.540\pm0.001$ &$0.541\pm0.002$ &$0.545\pm0.001$ &$0.545\pm0.003$ \\ \bottomrule
\end{tabular}
\caption{\Revision{Performance comparison between saliency prediction models.}}
\label{tab:compare_saliency_result}
\end{table}

\subsection{Ablation study}
To validate the modified multi-task learning scheme, ablation study is performed. The multi-task learning schemes following \equationref{eq:loss_1,eq:loss_2,eq:loss_3} are compared, and they are denoted as MTLS1, MTLS2, and MTLS3, respectively. Please note that the best-performing MTLS3 is used for benchmark comparison in \sectionref{sec:Benchmark comparison}. \figureref{fig:compare_scheme} in Appendix \ref{app:Performance evaluation} shows the training process for MTLS2 and MTLS3. With \figureref{fig:training,fig:compare_scheme}, we can see that overfitting occurs both for MTLS1 and MTLS2, but the overfitting is reduced in MTLS3. The training processes shown in \figureref{fig:training,fig:compare_scheme} are with optimized hyper-parameters. The resulting performances are compared in \tableref{tab:compare_scheme_result} in Appendix \ref{app:Performance evaluation}. We can see that MTLS3 outperforms the rest learning schemes both in classification and in saliency prediction.
\par
\Revision{
To validate the effects of using classification head that aggregates features from different depths, we create ablated versions of MT-UNet that use features from either the bottleneck or the top layer of the MT-UNet for classification, denoted as MT-UNetB and MT-UNetT, respectively. %\tableref{tab:compare_scheme_result} in \sectionref{app:Performance evaluation} shows the results.
Results are presented in \tableref{tab:compare_scheme_result} in Appendix \ref{app:Performance evaluation}.
We can see that MT-UNet generally performs better than MT-UNetT and MT-UNetB.}

\section{Discussion}
In this paper, we build the MT-UNet model and propose a further optimized multi-tasking learning scheme for saliency prediction and disease classification with CXR images. While a multi-task learning model has the potential of enhancing the performances for all component tasks, a proper training scheme is one of the key factors to fully unveil its potentiality. As shown in \tableref{tab:compare_scheme_result}, MT-UNet with the standard multi-task learning scheme may barely outperform existing models for saliency prediction or image classification.
\par
\Revision{
Several future work could be done to improve this study. The first would be the expansion of the gaze tracking dataset for medical images. So far, only $1083$ CXR images are publicly available with radiologist's gaze behavior, limiting extensive studies of gaze-tracking assisted machine learning methods in the medical field.
Also, more dedicated studies on multi-task learning methods, especially for small datasets, can be helpful for medical machine learning tasks. Overfitting and data deficiency are the lingering challenges encountered by many studies. A better multi-task learning method may handle these challenges more easily.}

% Acknowledgments---Will not appear in anonymized version
\midlacknowledgments{We would like to thank physionet.org for providing the open platform for dataset sharing, and we also like to express our gratitude to contributors who collected, organised and published the multi-modal chest X-ray dataset for this research. This research is supported by Compute Canada and the Natural Sciences and Engineering Research Council of Canada (NSERC).}

\bibliography{midl-fullpaper}

\appendix

\section{\Revision{Mathmatical deriviation of vicious circle for overfitting}}
\label{app:math}
Let $L\geq 0$ be the loss for a task, $\mathcal{T}$, and $\sigma>0$ be the variance estimator for $L$ used in \equationref{eq:loss_1}. Therefore, the loss for $\mathcal{T}$ following \equationref{eq:loss_1} can be expressed as:
\begin{equation}
    \mathcal{L} = \frac{L}{\sigma^2}+\ln(\sigma+1).
\end{equation}
The partial derivative of $\mathcal{L}$ with respect to $\sigma$ is:
\begin{equation}
    \frac{\partial \mathcal{L}}{\partial \sigma} = -\frac{2L}{\sigma^3}+\frac{1}{\sigma+1}.
\end{equation}
%During a gradient based optimization process, $\sigma$ achieves equilibrium (remains unchanged after gradient descend) when $\frac{\partial \mathcal{L}}{\partial \sigma}=0$. Therefore, during training, given $L$, $\sigma$ converges to the equilibrium value:
%During a gradient based optimization process, to minimize $\mathcal{L}$, $\sigma$ converges to the equilibrium value when $\frac{\partial \mathcal{L}}{\partial \sigma}=0$. Therefore, given $L$, the equilibrium value for $\sigma$ is:
%\begin{equation}
%    \tilde{\sigma} = \sqrt{\frac{L+\sqrt{L(L+4)}}{2}}\mbox{, s.t. }\left.\frac{\partial \mathcal{L}}{\partial \sigma}\right|_{\sigma=\tilde{\sigma}} = 0
%\label{eq:sigma tilde}
%\end{equation}
%which is a monotonically increasing function with respect to $L$.
%Therefore, when $L$ decreases due to overfitting, with \equationref{eq:sigma tilde}, we know that $\sigma$ will also decrease. The decreased $\sigma$ leads to an increase in the effective learning rate for $\mathcal{T}$, forming a vicious circle of overfitting.
During a gradient based optimization process, to minimize $\mathcal{L}$, $\sigma$ converges to the equilibrium value ($\sigma$ remains unchanged after gradient descend) which is achieved when $\frac{\partial \mathcal{L}}{\partial \sigma}=0$. Therefore, the following equation holds when $\sigma$ is at its equilibrium value, denoted as $\tilde{\sigma}$:
\begin{equation}
    L = \frac{\tilde{\sigma}^3}{2\tilde{\sigma}+2}
\end{equation}
which is calculated by letting $\frac{\partial \mathcal{L}}{\partial \sigma}=0$. Let $f(\tilde{\sigma}) = L$, $\tilde{\sigma}>0$, we can calculate that:
\begin{equation}
    \frac{d f(\tilde{\sigma})}{d \tilde{\sigma}} = \frac{\tilde{\sigma}^2(2\tilde{\sigma} + 3)}{2(\tilde{\sigma} +1)^2} >0, \quad \forall \tilde{\sigma}>0.
\end{equation}
Therefore, we know that $f(\tilde{\sigma})$ is strictly monotonically increasing with respect to $\tilde{\sigma}$, and hence the inverse function of $f(\tilde{\sigma})$, $f^{-1}(\cdot)$, exists. More specifically, we have:
\begin{equation}
    \tilde{\sigma} = f^{-1}(L).
\end{equation}
As a pair of inverse functions share the same monotonicity, we know that $\tilde{\sigma} = f^{-1}(L)$ is also strictly monotonically increasing. Thus, when $L$ decreases due to overfitting, we know that $\tilde{\sigma}$ will decrease accordingly, forcing $\sigma$ to decrease. The decreased $\sigma$ leads to an increase in the effective learning rate for $\mathcal{T}$, forming a vicious circle of overfitting.

\section{Training settings}
\label{app:Training settings}
\Revision{We use the Adam optimizer with default parameters \cite{kingma2014adam} and the RLRP scheduler for all the training processes. The RLRP scheduler reduces $90\%$ of the learning rate when validation loss stops improving for $P$ consecutive epochs, and reset model parameters to an earlier epoch when the network achieves the best validation loss. All training and testing are performed with the PyTorch framework \cite{paszke2019pytorch}. Hyper-parameters for optimizations are learning rate $r$, and $P$ in RLRP scheduler. The dataset is randomly partitioned into $70\%$, $10\%$ and $20\%$ subsections for training, validation and testing, respectively. The random data partitioning process preserves the balanced dataset characteristic, and all classes have equal share in all sub-datasets. All the results presented in this paper are based on at least $5$ independent trainings with same hyper-parameters. NVIDIA V100 and A100 GPUs (Santa Clara, USA) were used.}

\section{Performance evaluation}
\label{app:Performance evaluation}

\begin{table}[h]
\centering
\setlength{\tabcolsep}{3pt}
\begin{tabular}{@{}cccccc@{}}
\toprule
Metrics &MTLS1 &MTLS2 &MTLS3 &MT-UNetB &MT-UNetT \\ \midrule
KLD $\downarrow$ &$0.730\pm0.007$ &$0.738\pm0.006$ &$\bm{0.726}\pm0.004$ &$0.730\pm0.003$ &$0.734\pm0.007$ \\
CC $\uparrow$ &$0.566\pm0.005$ &$0.563\pm0.005$ &$\bm{0.569}\pm0.004$ &$0.568\pm0.003$ &$0.561\pm0.007$ \\
HS $\uparrow$ &$0.547\pm0.002$ &$0.545\pm0.002$ &$\bm{0.548}\pm0.001$ &$\bm{0.548}\pm0.001$ &$0.544\pm0.003$ \\
ACC $\uparrow$ &$0.649\pm0.041$ &$0.638\pm0.019$ &$\bm{0.670}\pm0.018$ &$0.653\pm0.013$ &$0.649\pm0.011$ \\
AUC $\uparrow$ &$0.832\pm0.019$ &$0.832\pm0.010$ &$0.843\pm0.012$ &$0.836\pm0.009$ &$\bm{0.847}\pm0.008$ \\
AUC-Y1 $\uparrow$ &$0.859\pm0.014$ &$0.861\pm0.015$ &$0.864\pm0.014$ &$0.859\pm0.007$ &$\bm{0.883}\pm0.005$ \\
AUC-Y2 $\uparrow$ &$0.906\pm0.016$ &$0.913\pm0.005$ &$\bm{0.912}\pm0.008$ &$0.907\pm0.011$ &$0.910\pm0.006$ \\
AUC-Y3 $\uparrow$ &$0.682\pm0.035$ &$0.672\pm0.010$ &$\bm{0.711}\pm0.027$ &$0.694\pm0.023$ &$0.695\pm0.025$ \\
\bottomrule
\end{tabular}
\caption{\Revision{Ablation study performance comparison.}}
\label{tab:compare_scheme_result}
\end{table}

\begin{figure}[thbp]
    \centering
    \subfigure[MTLS2 losses]{
	   \includegraphics[width=0.22\textwidth]{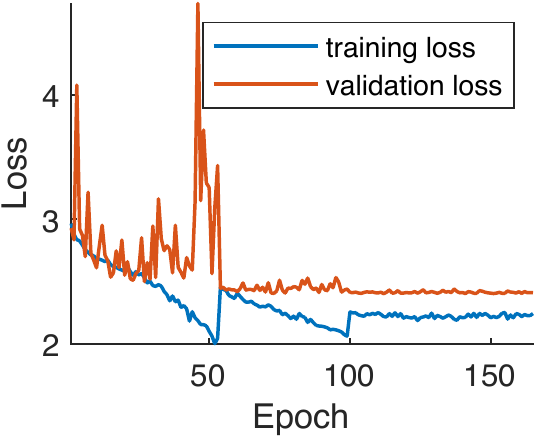}
	   \label{fig:MT-UNet-2 losses}
	   }
    \hspace{0.01em}
    \subfigure[MTLS2 $\sigma$]{
	   \includegraphics[width=0.22\textwidth]{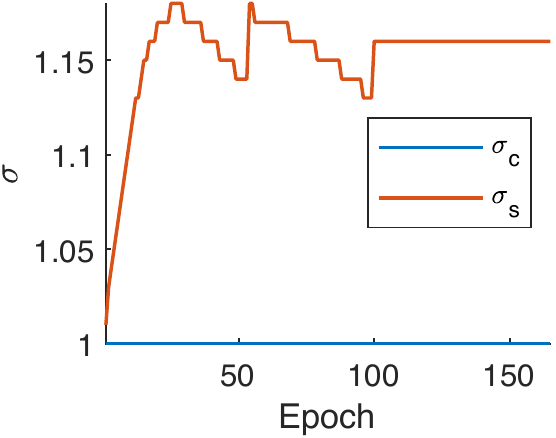}
	   \label{fig:MT-UNet-2 sigma}
	   }
	\hspace{0.01em}
	\subfigure[MTLS3 losses]{
	   \includegraphics[width=0.22\textwidth]{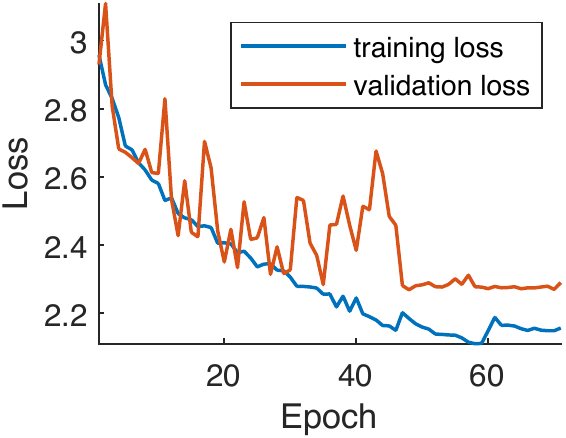}
	   \label{fig:MT-UNet-3 losses}
	   }
    \hspace{0.01em}
    \subfigure[MTLS3 $\sigma$]{
	   \includegraphics[width=0.22\textwidth]{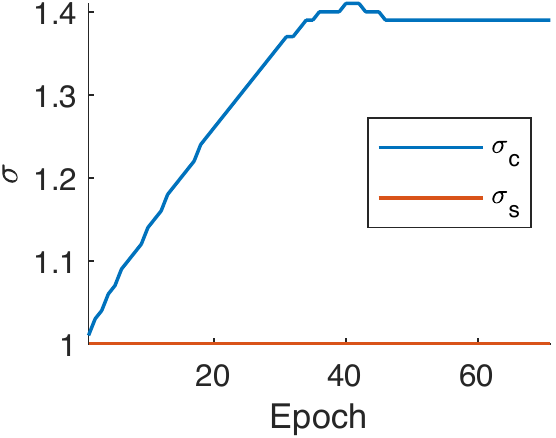}
	   \label{fig:MT-UNet-3 sigma}
	   }
\caption{Multi-task learning schemes comparison}
\label{fig:compare_scheme}
\end{figure}

\setlength{\tabcolsep}{3.5pt}
\begin{table}[h]
\centering
\begin{tabular}{@{}c|ccc@{}}
\toprule
   & Normal               & Enlarged heart & Pneumonia \\ \midrule
\begin{turn}{90} \parbox{2.0cm}{\centering Truth}  \end{turn}
&
\includegraphics[width = \figwidth\textwidth]{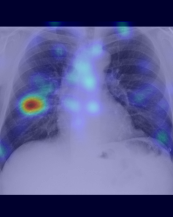}\hspace{\figgap pt}
\includegraphics[width = \figwidth\textwidth]{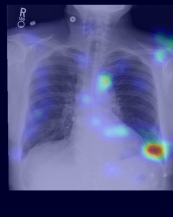}\hspace{\figgap pt}
\includegraphics[width = \figwidth\textwidth]{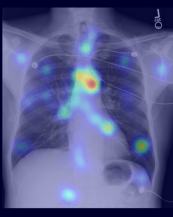}\hspace{\figgap pt}
&  
\includegraphics[width = \figwidth\textwidth]{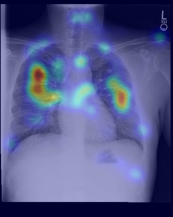}\hspace{\figgap pt}
\includegraphics[width = \figwidth\textwidth]{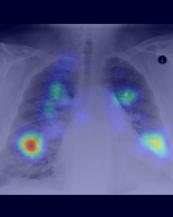}\hspace{\figgap pt}
\includegraphics[width = \figwidth\textwidth]{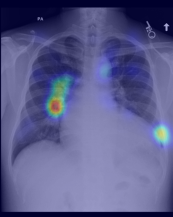}\hspace{\figgap pt}
&
\includegraphics[width = \figwidth\textwidth]{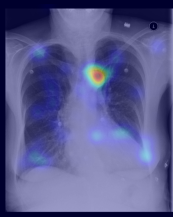}\hspace{\figgap pt}
\includegraphics[width = \figwidth\textwidth]{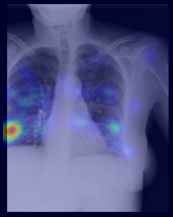}\hspace{\figgap pt}
\includegraphics[width = \figwidth\textwidth]{Saliency/Y1_37_truth.png}\hspace{\figgap pt}
\\
\begin{turn}{90} \parbox{2.0cm}{\centering MT-UNet}  \end{turn} 
&
\includegraphics[width = \figwidth\textwidth]{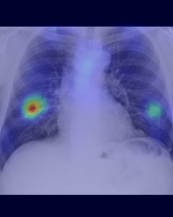}\hspace{\figgap pt}
\includegraphics[width = \figwidth\textwidth]{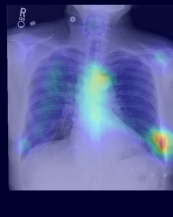}\hspace{\figgap pt}
\includegraphics[width = \figwidth\textwidth]{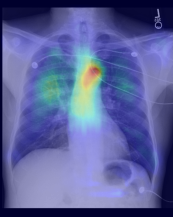}\hspace{\figgap pt}
&  
\includegraphics[width = \figwidth\textwidth]{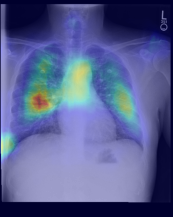}\hspace{\figgap pt}
\includegraphics[width = \figwidth\textwidth]{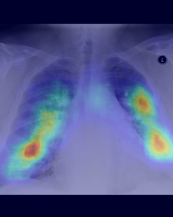}\hspace{\figgap pt}
\includegraphics[width = \figwidth\textwidth]{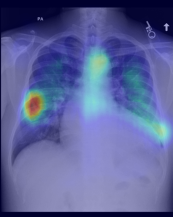}\hspace{\figgap pt}
&
\includegraphics[width = \figwidth\textwidth]{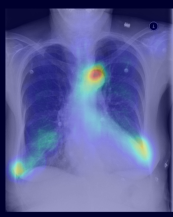}\hspace{\figgap pt}
\includegraphics[width = \figwidth\textwidth]{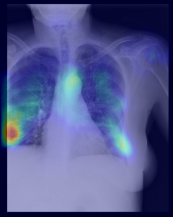}\hspace{\figgap pt}
\includegraphics[width = \figwidth\textwidth]{Saliency/Y1_37_MT-UNet.png}\hspace{\figgap pt}
\\
\begin{turn}{90} \parbox{2.0cm}{\centering UNet}  \end{turn}
&
\includegraphics[width = \figwidth\textwidth]{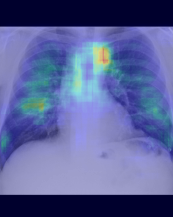}\hspace{\figgap pt}
\includegraphics[width = \figwidth\textwidth]{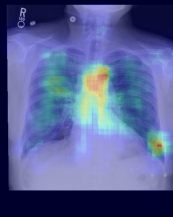}\hspace{\figgap pt}
\includegraphics[width = \figwidth\textwidth]{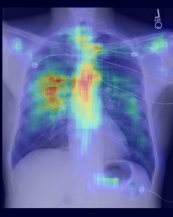}\hspace{\figgap pt}
&  
\includegraphics[width = \figwidth\textwidth]{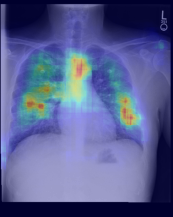}\hspace{\figgap pt}
\includegraphics[width = \figwidth\textwidth]{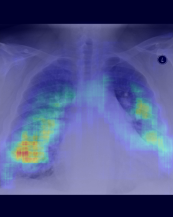}\hspace{\figgap pt}
\includegraphics[width = \figwidth\textwidth]{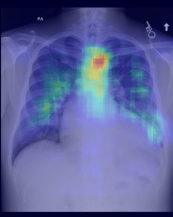}\hspace{\figgap pt}
&
\includegraphics[width = \figwidth\textwidth]{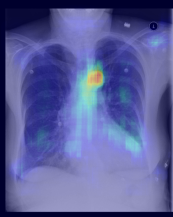}\hspace{\figgap pt}
\includegraphics[width = \figwidth\textwidth]{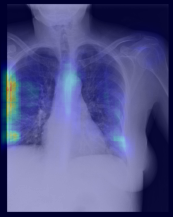}\hspace{\figgap pt}
\includegraphics[width = \figwidth\textwidth]{Saliency/Y1_37_UNet-SaliencyOnly.png}\hspace{\figgap pt}
\\
\begin{turn}{90} \parbox{2.0cm}{\centering SimpleNet}  \end{turn}
&
\includegraphics[width = \figwidth\textwidth]{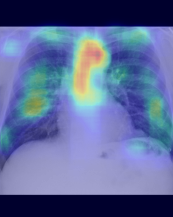}\hspace{\figgap pt}
\includegraphics[width = \figwidth\textwidth]{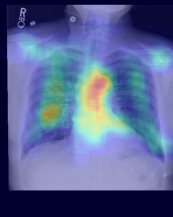}\hspace{\figgap pt}
\includegraphics[width = \figwidth\textwidth]{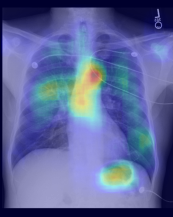}\hspace{\figgap pt}
&  
\includegraphics[width = \figwidth\textwidth]{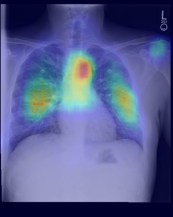}\hspace{\figgap pt}
\includegraphics[width = \figwidth\textwidth]{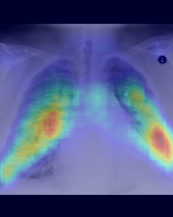}\hspace{\figgap pt}
\includegraphics[width = \figwidth\textwidth]{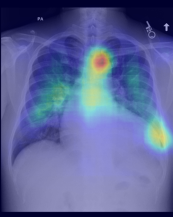}\hspace{\figgap pt}
&
\includegraphics[width = \figwidth\textwidth]{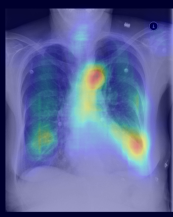}\hspace{\figgap pt}
\includegraphics[width = \figwidth\textwidth]{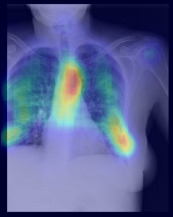}\hspace{\figgap pt}
\includegraphics[width = \figwidth\textwidth]{Saliency/Y1_37_SimpleNet.png}\hspace{\figgap pt}
\\
\begin{turn}{90} \parbox{2.0cm}{\centering MSINet}  \end{turn}
&
\includegraphics[width = \figwidth\textwidth]{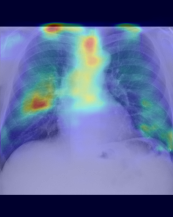}\hspace{\figgap pt}
\includegraphics[width = \figwidth\textwidth]{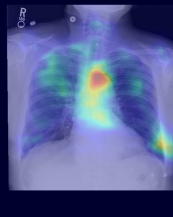}\hspace{\figgap pt}
\includegraphics[width = \figwidth\textwidth]{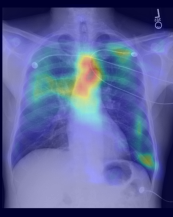}\hspace{\figgap pt}
&  
\includegraphics[width = \figwidth\textwidth]{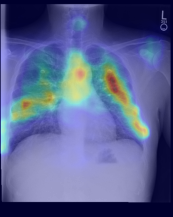}\hspace{\figgap pt}
\includegraphics[width = \figwidth\textwidth]{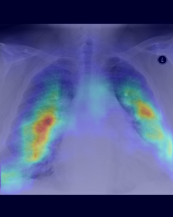}\hspace{\figgap pt}
\includegraphics[width = \figwidth\textwidth]{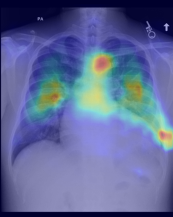}\hspace{\figgap pt}
&
\includegraphics[width = \figwidth\textwidth]{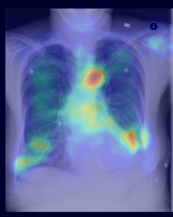}\hspace{\figgap pt}
\includegraphics[width = \figwidth\textwidth]{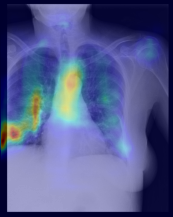}\hspace{\figgap pt}
\includegraphics[width = \figwidth\textwidth]{Saliency/Y1_37_MSINet.png}\hspace{\figgap pt}
\\
\begin{turn}{90} \parbox{2.0cm}{\centering VGGSSM}  \end{turn}
&
\includegraphics[width = \figwidth\textwidth]{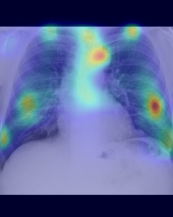}\hspace{\figgap pt}
\includegraphics[width = \figwidth\textwidth]{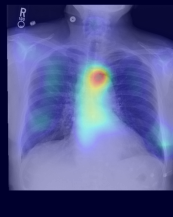}\hspace{\figgap pt}
\includegraphics[width = \figwidth\textwidth]{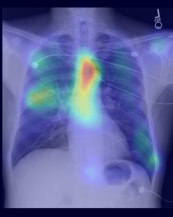}\hspace{\figgap pt}
&  
\includegraphics[width = \figwidth\textwidth]{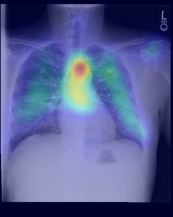}\hspace{\figgap pt}
\includegraphics[width = \figwidth\textwidth]{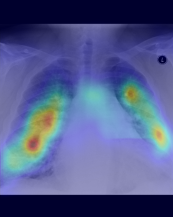}\hspace{\figgap pt}
\includegraphics[width = \figwidth\textwidth]{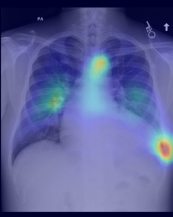}\hspace{\figgap pt}
&
\includegraphics[width = \figwidth\textwidth]{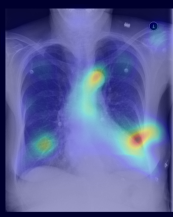}\hspace{\figgap pt}
\includegraphics[width = \figwidth\textwidth]{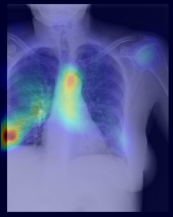}\hspace{\figgap pt}
\includegraphics[width = \figwidth\textwidth]{Saliency/Y1_37_VGGSSM.png}\hspace{\figgap pt}
\\
\bottomrule
\end{tabular}
\caption{Visualization of predicted saliency distributions. The ground truth and predicted saliency distributions are overlaid over CXR images. Jet colormap is used for saliency distributions where warmer (red and yellow) colors indicate higher concentration of saliency and colder (green and blue) colors indicate lower concentration of saliency.}
\label{tab:cam visual}
\end{table}

\end{document}